\documentstyle[aps,prb,multicol,epsf,amssymb]{revtex}

\def\gs{\gtrsim}
\def\ls{\lesssim}
\def\gdot{\dot{\gamma}}
\def\be{\begin{equation}}
\def\en{\end{equation}}                  
\newcommand{\bi}[1]{\mbox{\boldmath$#1$}}
\newcommand{\av}[1]{\langle{#1}\rangle}

\def\q{{\footnotesize{\it q}}\kern -5pt {\footnotesize{\it q}}}
\def\k{{\footnotesize{\it k}}\kern -5pt {\footnotesize{\it k}}}
\def\seq{\sim \kern -12pt \lower 5pt \hbox{$\displaystyle =$}}
\def\nnabla{\nabla\kern-3.3mm\nabla}

\def\ge{> \kern -12pt \lower 5pt \hbox{$\displaystyle =$}}
\def\le{< \kern -12pt \lower 5pt \hbox{$\displaystyle =$}}

\def\ep{\epsilon}

\def\bea{\begin{eqnarray}}
\def\ena{\end{eqnarray}}

\begin{document}
\draft
\bibliographystyle{prsty}
\title{Dynamics and Rheology of a Supercooled Polymer Melt in Shear Flow}
\author{Ryoichi Yamamoto and Akira Onuki}
\address{Department of Physics, Kyoto University, Kyoto 606-8502, Japan}
\date{\today}
\maketitle

\begin{abstract}
Using molecular 
dynamics simulations, we study 
dynamics of  a model polymer melt composed 
of short chains with  bead number  $N=10$ in supercooled states. 
In  quiescent conditions,   the stress relaxation function 
 $G(t)$  is calculated,  which exhibits  a stretched 
exponential relaxation  on the time scale of   
the $\alpha$ relaxation time  $\tau_\alpha$ 
and ultimately follows 
the Rouse dynamics characterized by the time  
$\tau_{\rm R} \sim N^2 \tau_\alpha$.  
After application of shear $\gdot$,  
transient stress growth $\sigma_{xy}(t)/\gdot$ first 
obeys the linear growth  $\int_0^t dt'G(t')$  
for strain less than 0.1 but  
 saturates into a non-Newtonian viscosity for larger strain.
In steady states,  shear-thinning and elongation of chains 
into ellipsoidal shapes take place 
 for  shear $\gdot$ larger than $\tau_{\rm R}^{-1}$.
 In such strong shear, we find that the  chains 
undergo random tumbling motion taking stretched and 
compact shapes alternatively.  
We  examine the validity of the stress-optical 
 relation between the  anisotropic parts of 
 the stress tensor and the dielectric tensor, 
which are violated in transient states due to the presence of  
a large glassy component of the stress.
We furthermore  introduce  time-correlation 
functions in shear to  calculate  the 
shear-dependent relaxation times, $\tau_\alpha (T,\gdot)$ 
and  $\tau_{\rm R} (T,\gdot)$, which decrease nonlinearly 
as functions of $\gdot$ 
in the shear-thinning regime.  
\end{abstract}


\begin{multicols}{2}

\section{Introduction}

The dynamics and rheology of glassy polymers 
are known to be very complicated 
and are still not well understood. 
We summarize salient features of such systems below.

First, in  the linear response regime, 
 thermal relaxations  
 of the chain conformations occur 
  from microscopic 
 to macroscopic time scales, as revealed in 
 measurements of  stress and dielectric 
 responses.\cite{matsuoka,stroble} 
In a relatively early stage, 
the stress relaxation function 
$G(t)$, which describes linear response to 
small shear deformations, 
 can be fitted to  the  
 Kohlrausch-Williams-Watts (KWW) form, 
\be 
G_{\rm G}(t) =  G_0 \exp [- (t/\tau_{\rm s})^c ],   
\label{eq:1.1}
\en 
after a  microscopic transient 
time $t_{\rm tra}$.  The time $\tau_{\rm s} (\gg \tau_{\rm tra})$ 
is of the order of the structural $\alpha$ 
relaxation time $\tau_\alpha$ (to be defined in (3.8) below), which 
grows dramatically as the temperature $T$ is 
lowered towards the glass transition 
temperature $T_{\rm g}$.  
When  the time $t$ considerably exceeds $\tau_{\rm s}$,  the relaxation of 
the chain conformations is relevant and 
is well described by the Rouse or reptation dynamics,  
depending on $N<N_{\rm e}$ or $N>N_{\rm e}$, respectively.\cite{doi} 
Here $N$ is the polymerization index and 
$N_{\rm e}$ is that between entanglements on a chain. 
For short chain systems with  $N<N_{\rm e}$,    
the overall behavior in the time region 
 $t\gg t_{\rm tra}$ may be  expressed  as 
\be 
G(t) =   G_{\rm G}(t)+ G_{\rm R}(t).  
\label{eq:1.2}
\en 
The $G_{\rm R}(t)$ is 
the stress time-correlation  function  in the 
Rouse model whose terminal relaxation time 
$\tau_{\rm R}$ is of order $N^2 \tau_\alpha$. 
For entangled  chain systems with $N>N_{\rm e}$,  
the KWW function in (1.1) 
 is followed by the power-law decay,  
\be 
G(t) \cong e^{-1} G_0 (t/\tau_{\rm s} )^{-\nu}
\label{eq:1.3}
\en   
with $\nu \sim 0.5$
until the rubbery plateau $G(t)\simeq G_N^{(0)}$ 
is reached,\cite{matsuoka,stroble} 
where $G_N^{(0)}$ assumes the modulus $nk_{\rm B}T/N_{\rm e}$ 
of entangled polymers with $n$ being the bead number density. 
Ultimately, $G(t)$ follows  the  reptation 
relaxation $G_{\rm rep}(t)$ 
on the time scale of a very long  reptation time 
$\tau_{\rm rep}$.  These  hierarchical relaxations arise from 
 rearrangements  of  jammed 
  atomic configurations 
 and  subsequent evolution 
 of chain conformations.  
They also give rise to 
the corresponding characteristic 
behaviors in the frequency-dependent 
shear modulus $G^*(\omega)\equiv  i\omega 
\int_0^\infty dt e^{-i\omega t}G(t)$, 
depending on the frequency $\omega$ relative to 
the inverse characteristic times introduced so far.  
\cite{matsuoka,stroble,doi}

Second, in the nonlinear 
 response regime, glassy fluids 
generally exhibit highly viscous non-Newtonian flow 
close to (but above) $T_{\rm g}$ even if they are 
low-molecular-weight fluids.\cite{Simmons} 
In such fluids, if $\gdot > \tau_\alpha^{-1}$,     
atomic rearrangements are 
induced not by thermal agitations 
but by externally  applied shear.\cite{yo1,yo2,yo3} 
  In chain systems 
 without entanglements, on one hand,  
shear thinning occurs at sufficiently 
high (but sometimes unrealistically large) shear rates 
due to chain elongation.\cite{KLH,ch,khare,Moore,Cummings}
In entangled polymers, on the other hand, 
shear thinning occurs at 
very small shear larger than $\tau_{\rm rep}^{-1}$,
where  disentanglements are induced by shear.\cite{doi}
Thus supercooled chain systems are most 
easily driven into a nonlinear response 
regime even by extremely small shear, though 
the crossover shear stress from linear to nonlinear regimes 
 may not be very small. 
Furthermore, in glassy fluids below  $T_{\rm g}$, 
plastic deformations are often 
induced in the form of large-scale shear bands 
above a yield stress (corresponding to a few $\%$ strain). 
\cite{Ara1,Ara2}  It is  of great importance 
to understand  how these nonlinear 
effects  occur depending  on  $\gdot$, 
$T$, and $N$.

Third,  in rheological experiments 
on polymers,  use has been made of the 
stress-optical relation between the deviatoric 
(anisotropic) parts of the dielectric tensor $\ep_{\alpha\beta}$ 
(at optical frequencies) and the average stress tensor 
$\sigma_{\alpha\beta}$.\cite{doi,onuki-doi}  
In shear flow  with 
mean velocity $\gdot y$ in the $x$ direction,  
 it is expressed as 
\be 
\ep_{xy}= C_0 \sigma_{xy},\quad 
\ep_{\alpha\alpha}- \ep_{\beta\beta}= 
C_0 (\sigma_{\alpha\alpha}- \sigma_{\beta\beta}),  
\label{eq:1.4}
\en 
where $C_0$ is called the stress-optical coefficient.
For melts this relation excellently holds  at 
relatively high $T$ ($> T_{\rm g}$) 
 for general time-dependent 
nonlinear shear deformations.
If  measurements are made in 
steady states, it holds even close to $T_{\rm g}$.\cite{muller}    
For its validity, we need to 
require that  the form contribution  to  
$\ep_{\alpha\beta}$ is  negligible as  compared to the intrinsic 
contribution\cite{onuki-doi} 
and that the glassy part 
of  the stress is negligible as compared to the 
 usual entropic part.  
Thus  it   is  violated  when 
the form part is relevant such as in 
  polymer solutions 
close to the  demixing critical point   
or when  measurements are made in transient states close to 
$T_{\rm g}$.  In the latter case, 
as is evident  from  the enhancement of
$\tau_{\rm s}$ in (1.1),  the glassy part 
 of the stress is dominant  
 for  relatively rapid deformations.\cite{muller,muller1,inoue,matsuyama}

While the predictive  power of analytic theories 
in polymer science is still poor, 
computer simulations\cite{kremer,carmesin,deutsch}
can provide us a useful tool 
to investigate 
the microscopic origins of experimentally observed 
macroscopic phenomena.  
In  quiescent states, diffusive 
motions in supercooled melts have been extensively 
studied using 
molecular dynamics (MD)\cite{Roe,bennemann,baschnagel,moe,zon}
and Monte Carlo\cite{okun,paul,binder} simulations.
In another  application, 
nonequilibrium molecular dynamics (NEMD)  
simulations have been useful to investigate 
chain deformations and rheology 
in flow.\cite{KLH,ch,khare,Moore,Cummings,muller1,rouault}  
In particular,  
Kr\"oger {\it et al.}\cite{muller1}  
studied  the molecular mechanisms 
of the violations of stress-optical behavior 
for a melt consisting $M=260$ chains 
with bead number  $N=30$  after 
application of elongational flow.

In this paper, we will present results of  
very long MD simulations 
to study  linear and nonlinear dynamics 
of a supercooled polymer melt 
in the absence and presence of shear flow. 
Long simulation times are needed to calculate 
the terminal relaxation of $G(t)$, which 
has not yet been undertaken in the literature. 
As a new finding, we will show that 
each chain in our melt system is  changing 
its orientation (tumbling) 
randomly in shear flow. 
Use will be made of techniques and 
concepts introduced in  our previous 
papers  on supercooled binary mixtures 
under  shear flow.\cite{yo1,yo2,yo3} Some 
 of  our results were  already published 
elsewhere.\cite{yo3,Onukibook}

\setcounter{equation}{0}
\section{Model and simulation method}

Our system is composed 
of $M=100$ chains with $N=10$ beads  
confined in  a cubic box 
with length $L=10\sigma$ 
and volume $V=L^3=10^3\sigma^3$. 
The number density is fixed at  
$n= NM/V=1/\sigma^3$, which results in 
severely jammed configurations at low $T$.  
All the bead particles interact with a truncated  
Lennard-Jones potential defined by\cite{kremer} 
\begin{equation}
U_{\rm LJ}(r)=4\epsilon \left 
[\left(\frac{\sigma}{r}\right)^{12}-\left(\frac{\sigma}{r}
\right)^{6}\right]+\epsilon \quad (r<2^{1/6}\sigma). 
\label{eq:2.1}
\end{equation} 
The right hand side  
is minimum at  $r=2^{1/6}\sigma$ 
and the potential is truncated for larger $r$ ($U_{\rm LJ}(r)=0$ for 
 $r>2^{1/6}\sigma$). By using   the  repulsive part of the 
Lennard-Jones potential only, 
 we may prevent spatial overlap of the particles.\cite{kremer}  
Consecutive beads on each chain are connected by  
an anharmonic spring of the form, 
\begin{equation}
U_{\rm F}(r)= -\frac{1}{2}k_cR_0^2 \ln 
[1-({r}/{R_0})^2]
\label{eq:2.2}
\end{equation}
with $k_c=30\epsilon/\sigma^2$ and $R_0=1.5\sigma$. 
In our simulation  the bond lengths  
$b_{j}^k \equiv  |{\bi R}_{j}^k-{\bi R}_{j+1}^k|$ 
($1 \leq j \leq N-1)$ 
between consecutive beads on the same chain $k$  were 
 very close to the minimum distance 
$b_{\rm min} \cong  0.96\sigma$ of the sum 
$U_{\rm LJ}(r)+U_{\rm F}(r)$. The  deviations $b_j^k-b_{\rm min}$  
were only on the order of a few $\%$ of $b_{\rm min}$ 
 for any $T$ and $\gdot$  realized in our study.

Microscopic expressions for physical quantities 
can be expressed  in terms 
of the momentum and position vectors of the $j$-th bead on 
the $k$-th chain, 
 ${\bi R}_j^k$ and ${\bi p}_j^k$,  where $j=1, \cdots, N$ and  
 $k=1,\cdots, M$.    For example, 
the space integral of the microscopic stress tensor reads 
\bea 
{\Pi}_{\alpha \beta}^{\rm T}(t)&=& \frac{1}{m} 
\sum_{k=1}^{M}\sum_{j=1}^N  p_{j \alpha}^k p_{j \beta}^k
- \sum_{\rm all~ pairs} U_{\rm LJ}'(\xi)
\frac{\xi_\alpha \xi_\beta}{\xi}  \nonumber \\
&&- \sum_{k=1}^M \sum_{j=1}^{N-1} U_{\rm F}'(\xi)
\frac{\xi_{\alpha}\xi_{\beta}}{\xi} , 
\label{eq:2.3} 
\ena  
where $m$ is  the mass of a bead, 
$U_{\rm LJ}'(\xi)= dU_{\rm LJ}(\xi)/d\xi$,  and 
$U_{\rm F}'(\xi)= dU_{\rm F}(\xi)/d\xi$. 
Here ${\bi \xi}= (\xi_x,\xi_y,\xi_z)$ 
in the right hand side 
represents the relative vector 
${\bi R}_j^k- {\bi R}_{j'}^{k'}$ 
 between the two beads,
  ${\bi R}_j^k$ and ${\bi R}_{j'}^{k'}$,  
 in the second term and 
  the relative vector ${\bi R}_j^k- {\bi R}_{j+1}^{k}$ 
 between the two consecutive 
 beads,  ${\bi R}_j^k$ and ${\bi R}_{j+1}^{k}$,
  of  the same chain in the third term.  
To avoid cumbersome notation, 
we will write the bead positions  simply  as 
${\bi R}_j$ $(j=1, \cdots, N)$  suppressing 
the index $k$.  When they will appear in 
the  statistical averages 
$\av{\cdots}$,  the average over  all the chains 
$\sum_{k=1}^M(\cdots)/M$ will be implied 
even if not written explicitly.
Furthermore, it is convenient here 
to introduce the usual notation  
$ \sigma_{\alpha \beta}$ 
for the stress tensor by  
\be 
 \frac{1}{V} 
 {\Pi}_{\alpha \beta}^{\rm T} =  
p\delta_{\alpha \beta}- \sigma_{\alpha \beta}, 
\label{eq:2.4} 
\en 
where $p$ is the pressure  and  
the second  term is deviatoric.\cite{doi}  The 
$\sigma_{\alpha \beta}$ has 
already  appeared  
in the stress-optical relation (1.4).

We will hereafter 
measure space and time in units of $\sigma$ and  
$\tau_0\equiv ({m\sigma^{2}/\epsilon})^{1/2}$. 
The temperature $T$ will be measured in units of $\epsilon/k_{\rm B}$.
The original units will also be used when 
confusion may occur. 
Our simulations cover normal ($T=1.0$) and supercooled ($T=0.2$) 
states with and without shear flow ($\gdot=0$, $10^{-4}$, $10^{-3}$,
$10^{-2}$, and $10^{-1}$).
Simulation data were  taken after very long equilibration periods 
($\sim10^2\tau_R\simeq5\times10^6$ at $T=0.2$) so that 
 no appreciable aging (slow equilibration) effect 
 was detected in 
the course of taking data in various quantities such as the pressure 
or the density time-correlation functions.
(i) In quiescent cases,  we impose the micro-canonical condition 
and integrate  the  Newton's equations of motion, 
\be
\frac{d}{dt}{\bi R}_j  =\frac{1}{m} {\bi p}_j,  \quad 
\frac{d}{dt}{\bi p}_j  ={\bi f}_j, 
\label{eq:2.5}
\en
where ${\bi f}_j$ is the force
acting on the particle $j$ due to the potentials.  
Integration was performed with 
time increment $\Delta t = 0.005$ under the periodic boundary condition.  
Long time  simulations of order $10^2 \tau_{\rm R}$, which corresponds to 
$10^{9}$ MD steps for $T=0.2$,  were  performed.
In the previous simulations,\cite{kremer,Roe,bennemann,baschnagel,okun,paul,binder}
however, the integrated times did not much exceed 
$\tau_{\rm R}$ in supercooled states. 
(ii) In the presence of shear, 
rewriting the momentum deviations 
${\bi p}_j - m \gdot Y_j {\bi e}_x$ 
from the mean flow  as ${\bi p}_j$, we   
 integrated  the so-called 
SLLOD equations of motion,\cite{Allen,Evans} 
\bea 
\frac{d}{dt}{\bi R}_j
&=& \frac{1}{m}{\bi p}_j + \gdot Y_j{\bi e}_x, \nonumber\\
\frac{d}{dt}{\bi p}_j &=&
{\bi f}_j- \gdot p_{y_j}
{\bi e}_x-\hat{\zeta}{\bi p}_j, 
\label{eq:2.6}
\ena 
where ${\bi e}_x$ is the unit vectors in the $x$ (flow) direction, 
 ${\bi R}_j=(X_j,Y_j,Z_j)$, and ${\bi p}_j=(p_{xj},p_{yj},p_{zj})$. 
The  friction coefficient $\hat \zeta$ 
was set equal to 
\be
\hat{\zeta} =
{\sum_{j}({\bi f}_j\cdot{\bi p}_j-\gdot p_{xj} p_{yj})} \bigg / 
\sum_{j}{{\bi p}}_j^2.   
\label{eq:2.7}
\en
The temperature 
$T$ $(\equiv {\sum_{j}{{\bi p}}_j^2}/mNM$) 
could then be 
   kept at a desired value. 
The time increment was  $\Delta t=0.0025$.  
After an equilibration run in a quiescent state for $t < 0$, 
we gave all the particles  
the average flow velocity 
$\gdot Y_j{\bi e}_x$ 
at $t = 0$  and then 
imposed the Lee-Edwards boundary 
condition\cite{Allen,Evans} to  maintain the  shear flow.  
Steady sheared states were realized after 
transient relaxations.

\setcounter{equation}{0}
\section{Dynamics in quiescent states}

Although it is highly nontrivial, it has been confirmed by computer 
simulations\cite{kremer,Roe,bennemann,baschnagel,okun,paul,binder,kopf} 
that the single-chain near-equilibrium dynamics in unentangled melts 
can be reasonably well described by (or mapped onto) the simple Rouse 
model.
In the Rouse dynamics,
the relaxation time of the $p$-th  mode of a chain is 
expressed in terms of  a friction
coefficient  $\zeta$ and a segment length $b$ as\cite{verdier} 
\begin{equation}
\tau_p= {\zeta b^2}/[{12 k_{\rm B}T }\sin^{2}({\pi p}/{2N})] ,
\label{eq:3.1}
\end{equation}
where $1 \leq p\leq N-1$.
The Rouse relaxation time 
$\tau_{\rm R}$ is the slowest relaxation time, 
\begin{equation}
\tau_{\rm R} \equiv \tau_1 \cong N^2\zeta b^2/({3\pi^2k_{\rm B}T}) .
\label{eq:3.2}
\end{equation}
The  segment length $b$ in the corresponding  
Rouse model may be related to 
the variance of the end-to-end vector of a chain 
${\bi P}\equiv  {\bi R}_N - {\bi R}_1$ in our microscopic model by 
\begin{equation}
{\av{|{\bi P}|^2}}= b^2(N-1) . 
\label{eq:3.3}
\end{equation} 
 As a result,  $b$  is dependent on 
 $T$ but its dependence turns out to be weak as 
   $b=1.17$, $1.18,$ $1.19$ 
for $T= 1.0$, $0.4$, $0.2$, respectively.
Note that $b$ is  larger than the minimum distance 
$b_{\rm min}\cong 0.96$ of the bond-potential.  Let us consider  the  
 time-correlation function of ${\bi P}(t)$, 
 \be 
C(t)= {\av{{\bi P}(t+t_0)\cdot{\bi P}(t_0)}}/\av{|{\bi P}|^2},  
\label{eq:3.4}
\en 
which is normalized such that $C(0)=1$. 
Here $C(t)$ should be independent of 
the initial time $t_0$ in steady states in the limit of large 
system size. However, our system is not very large, 
so we took   the average over the initial time $t_0$. 
This statistical averaging  will not be mentioned  hereafter 
in showing our MD results of time-correlation 
functions.      In  the Rouse dynamics 
$C(t)$ is calculated as   
\begin{equation}
C_{\rm R}(t)   = 
\frac{2}{(N-1)N} 
\sum_{{\rm odd}~p} \cot^2 \bigg (
\frac{\pi p}{2N}\bigg ) 
e^{- {t}/{\tau_p}},
\label{eq:3.5}
\end{equation}
where the summation is over odd $p$ 
but the first term ($p=1$) is dominant 
in the whole time region (so we may determine 
$\tau_{\rm R}$ by $C(\tau_{\rm R})= e^{-1}$).  
As shown in Fig.1,  our MD data of 
$C(t)$ can  be excellently 
fitted to $C_{\rm R}(t)$. The  $\tau_{\rm R}$  
thus  determined 
 increases drastically with lowering $T$ as 
$\tau_{\rm R}=250$, $1800$, 
and $6\times 10^4$ for $T= 1.0$, $0.4$, and $0.2$, 
respectively. 
In the previous simulations on nonentangled polymer melts, 
\cite{bennemann,baschnagel,okun,paul,binder,kopf}
numerical results were 
consistent with the Rouse dynamics  
for small $p$ (large-scale motions), 
but deviations are enhanced for
large $p$ (small-scale motions)  in supercooled states.
Furthermore, 
we here give the expression for  the  stress relaxation 
function in the Rouse model,    
\be
G_{\rm R}(t)=\frac{n k_{\rm B} T }{N}\sum_{p=1}^{N-1}\exp(-{2t}/{\tau_p}),
\label{eq:3.6}
\en 
which is equal to $nk_{\rm B}T(N-1)/N$  at  $t =0$ and decays as 
$nk_{\rm B}TN^{-1} \exp(-2t/\tau_{\rm R})$ for $t \gs \tau_{\rm R}$.  
Since  $G(t)$ is much larger than 
$G_{\rm R}(t)$ in the relatively short time region $t<\tau_s$, 
they  can coincide only in the late stage.

In Fig.\ref{fsqt},  we 
 show the van Hove self-correlation function, 
\begin{equation}
F_q(t)= \frac{1}{N}\sum_{j=1}^N 
\av{\exp[i{\bi q}\cdot \Delta {\bi R}_j(t)]}, 
\label{eq:3.7}
\end{equation}
where $q=2\pi$,  
$\Delta {\bi R}_j(t)={\bi R}_j(t+t_0) 
-{\bi R}_j(t_0)$ is the  displacement 
vector of the $j$-th bead 
in the time interval $[t_0,t_0+t]$. 
The peak wave number of  the static structure 
 factor is given by $q\cong 2\pi$. 
We define the $\alpha$ relaxation time $\tau_\alpha$ 
from the condition,  
\be 
F_q(\tau_\alpha)=e^{-1} \quad (q=2\pi).  
\label{eq:3.8}
\en  
As has been reported in the literature, $\tau_\alpha$  increases 
drastically with lowering $T$.\cite{bennemann,baschnagel,moe,okun,paul}
In our case, we obtained 
$\tau_\alpha=0.91$, $5.8$, and $310$ 
for $T= 1.0$, $0.4$, and $0.2$, 
respectively. 
At $T=0.2$, where the particle 
motions  are considerably 
jammed, $F_q(t)$ 
exhibits a two-step relaxation 
and may be excellently fitted to 
the stretched exponential decay ($ 
\propto \exp [-(t/\tau_\alpha)^{0.64}]$)  
for $t \gs 10$. Thus our system  at $T=0.2$ 
has characteristic features of 
a supercooled state,  although its 
melting temperature is unknown.
We find $\tau_\alpha \sim 10^{-2} {\zeta b^2}/{ k_{\rm B}T}$ 
and $\tau_{\rm R} \sim N^2\tau_\alpha$ at any $T$. 
Particularly, for $T=0.2$,  we  obtain   
\begin{equation}
\tau_\alpha \cong  0.017 {\zeta b^2}/{ k_{\rm B}T}, 
\quad  
\tau_{\rm R} \cong 1.9 N^2\tau_\alpha. 
\label{eq:3.9}
\end{equation} 
The friction 
coefficient $\zeta$  in the mapped Rouse model 
grows  strongly as $T$ is 
lowered in supercooled states.

Now we discuss the linear viscoelastic behavior in 
supercooled states. In terms of 
$\Pi_{xy}^{\rm T}(t)$ in (2.3), the 
stress relaxation function $G(t)$ is written as\cite{Allen,Evans} 
\be
G(t) = \langle \Pi_{xy}^{\rm T}(t+t_0)
\Pi_{xy}^{\rm T}(t_0)\rangle /{k_{\rm B}TV} . 
\label{eq:3.10}
\en 
In Fig.\ref{gt}, 
we show  numerical data of $G(t)$, 
where  the average over the initial time 
$t_0$ was taken but the data become noisy 
at very large $t \sim \tau_{\rm R}$.  
In the very early stage  $t \ls 1$,  $G(t)$ 
oscillates  rapidly due to the vibrations of 
the bond vectors ${\bi b}_j= 
{\bi R}_j-{\bi R}_{j+1}$. 
 The initial value $G(0)$ takes a  large value ($\sim 100$ 
 in units of $\epsilon/\sigma^3$)   nearly independent 
 of $T$. For  $T=0.2$, $G(t)$ 
can be nicely fitted to the 
the stretched exponential  form (1.1) with $G_0 \cong  5$,  
$\tau_{\rm s}=90 \cong  0.33\tau_\alpha$, and $c=0.5$ in 
the time region  $1 \ls t \ls 10\tau_{\rm s}$. 
For $t\gs 50\tau_{\rm s}$ it approaches the Rouse stress relaxation function 
$G_{\rm R}(t)$ in (3.5).  The zero-frequency 
Newtonian viscosity is given by  
$\eta (0) = \int_0^\infty dt G(t)$,
 so it  consists of the glassy (monomeric) part, 
\be 
\eta_{\rm G}=\int_0^\infty dt G_{\rm G}(t)  \sim 10\tau_{\rm s}, 
\label{eq:3.11}
\en 
and the Rouse (polymeric) part,  
\be 
\eta_{\rm R}= \int_0^\infty dt G_{\rm R}(t)
\cong  0.808 TN^{-1}\tau_{\rm R}.
\label{eq:3.12}
\en 
The ratio $\eta_{\rm G}/\eta_{\rm R}$ is of order $1/TN$. 
They are of the same order in 
the present case of  $N=10$ and $T=0.2$. 
However,  we should have $\eta_{\rm G} \ll 
\eta_{\rm R}$  for much larger $N$.

To examine the orientation of the bonds,
we consider the orientational tensor, 
\be
Q_{\alpha\beta}(t) = \frac{1}{M}  \sum_{\rm chain}  \frac{1}{N-1}
\sum_{j=1}^{N-1} 
\frac{b_{j\alpha}}{b_0}\frac{b_{j\beta}}{b_0},  
\label{eq:3.13}
\en  
where $b_0^{-1} {\bi b}_j$ are 
the normalized bond vectors since 
 $|{\bi b}_j| \cong b_0$ as stated below (2.2). 
 Notice that in the Rouse model  
the space integral of the entropic stress tensor 
is given by the expression 
$\sigma^{\rm b}_{\alpha\beta} \equiv  
 {(3k_{\rm B}T b_0^2/b^2)}Q_{\alpha\beta}$, where $b$ 
 determined by  (3.3) appears 
 instead of $b_0$.  To compare  our microscopic  system 
and the simplified Rouse model,  
we calculated the time-correlation function, 
\begin{equation}
G_{\rm b}(t)=  \av{\sigma^{\rm b}_{xy}
(t+t_0)\sigma^{\rm b}_{xy}(t_0)}/{k_{\rm B}TV}, 
\label{eq:3.14}
\end{equation}
by integrating (2.5). 
As shown in Fig.\ref{grt}, 
 $G_{\rm b}(t)$ is fairly  close to  
 $G_{\rm R}(t)$ in (3.5) from the Rouse model.  
In particular, 
for $t\gs 0.1\tau_{\rm R}$, 
we find $G(t)\cong 
G_{\rm b}(t) \cong G_{\rm R}(t)$.

\setcounter{equation}{0}
\section{Steady state behavior in shear flow}

In Fig.\ref{vis}, we display the steady-state viscosity   
$\eta(\gdot)\equiv 
\sigma_{xy}/\gdot$ obtained at $T=0.2,~ 0.4,$ and  $1$, 
where the time average of the stress was taken.  
The crossover shear rate  
from  Newtonian to shear-thinning  behavior 
is given by $\tau_{\rm R}^{-1}\sim N^{-2}\tau_\alpha^{-1}$.  
We may introduce 
the Weisenberg number $Wi$ by 
\be 
Wi = \gdot \tau_{\rm R} \sim \gdot N^{2}\tau_\alpha.  
\label{eq:4.1}
\en 
In the non-Newtonian regime, we have $Wi>1$.
The shear stress at the crossover  
is of order $nk_{\rm B}T N^{-1}$, which is the elastic 
modulus of the Rouse model. 
The horizontal arrows indicate the linear Rouse 
viscosity $\eta_{\rm R}$ in (3.12), while the vertical arrows indicate
the points at which $\gdot=\tau_{\rm R}^{-1}$. 
In particular, the curve of  $T=0.2$ may be 
fitted to 
\begin{equation}
\eta (\gdot) \propto \gdot^{-\nu}
\label{eq:4.2}
\end{equation} 
with $\nu\cong 0.7$ for 
$\gdot\tau_{\rm R}\gs 1$. 
The   $\eta (\gdot)$  becomes insensitive to $T$ for 
very high shear ($Wi \gg 1$). 
In MD simulations of short chain systems 
in normal liquid states,\cite{ch,khare,Moore,Cummings} 
similar shear-thinning has been reported,  
where the crossover shear 
is  much higher, however.   
In MD simulations of supercooled  
simple binary mixtures,\cite{yo1,yo2,yo3}  
shear-thinning becomes apparent 
for $\gdot \gs \tau_\alpha^{-1}$, 
where $\tau_\alpha$ grows at low $T$ 
and which is consistent 
with (4.1) if we set  $N=1$.  

To demonstrate 
 the stress-optical law  in steady states,  
we show  steady-state data of $\sigma_{xy}/T$ vs $Q_{xy}$  
 for $T=0.2,0.4,$ and $1$ in Fig.6.
If the electric  polarization tensor of a bead  is uniaxial 
along the bond direction, the deviatoric part  of 
the dielectric tensor is proportional to that of 
the  tensor $Q_{\alpha\beta}$ in (3.13). 
In accord with the experiment,\cite{muller}  
our data collapse onto a universal curve independent 
of $T$ both in the linear  $(Q_{xy} \ls 0.05)$ 
and  nonlinear $(Q_{xy} \gs 0.05)$ regimes.

We  next consider  anisotropy of chain conformations in shear flow.
In Fig.\ref{ellips}(a),  
we plot the $x$-$y$ cross-section ($z=0$)  
of the steady state bead distribution function, 
\begin{equation}
g_{\rm s}({\bi r}) = \frac{1}{N} 
\sum_{j=1}^N\av{\delta({\bi R}_j-{\bi R}_{\rm G}-{\bf r})},
\label{eq:4.3}
\end{equation}
where  $\gdot=10^{-4}$,  $T=0.2$, and 
${\bi R}_{\rm G}={N}^{-1}\sum^{N}_{j=1}{\bi R}_j$ is the center of 
mass of a chain.
In Fig.\ref{ellips}(b), we also 
plot the  structure factor in the $q_x$-$q_y$ plane ($q_z=0$), 
\begin{equation}
S({\bi q})= \frac{1}{N^{2}} \sum_{i,j=1}^N 
\av{\exp[i{\bi q}\cdot({\bi R}_i -{\bi R}_j]} ,
\label{4.4}
\end{equation}
which  is proportional to the scattering intensity 
from labeled chains in shear.\cite{picot}  
 In these figures 
 $\theta$ is the relative angle 
of the ellipses with respect to 
the $y$ (shear gradient) direction. 
These figures demonstrate 
 high elongation of the chains 
 for $\gdot>\tau_{\rm R}^{-1}$.
As will be shown in Fig.\ref{aniso} below, 
they almost saturate into the shapes shown 
in Fig.\ref{ellips} once $\gdot$ exceeds $\tau_{\rm R}^{-1}$.

Let us define the tensor, 
\bea
I_{\alpha\beta} &=& \frac{1}{N^{2}}\sum_{i=1}^N\sum_{j=1}^N \av{
({R}_{i\alpha} -{R}_{j\alpha }) 
({ R}_{i\beta} -{R}_{j\beta})}\nonumber \\
&=& \frac{2}{N} 
\sum_{j=1}^N \av{
({R}_{j\alpha} -{R}_{{\rm G}\alpha }) 
({ R}_{j\beta} -{R}_{{\rm G}\beta})}. 
\label{4.5}
\ena 
For small ${\bi q}=(q_x,q_y,0)$, $S({\bi q})$ is  expanded as   
\bea
S({\bi q})&=& 1-\frac{1}{2}\sum_{\alpha,\beta=x,y}I_{\alpha\beta}
q_\alpha q_\beta +\cdots \nonumber\\
&=& 1-\frac{1}{2}a_1^2 ({\bi q}\cdot{\bi e}_1)^2 - 
\frac{1}{2}a_2^2 ({\bi q}\cdot{\bi e}_2)^2+\cdots, 
\label{4.6}
\ena 
where $\{{\bi e}_1, {\bi e}_2\}$ and  $\{a_1^2, a_2^2\}$ 
are the  unit eigenvectors and eigenvalues of the tensor 
$I_{\alpha\beta}$ ($\alpha,\beta\in x,y$).
The two lengths $a_1$ and $a_2$ correspond to the shorter 
and longer radii in the principal axes of the ellipses. 
 In terms of $\theta$,  we have 
 ${\bi e}_1= (-\sin\theta,\cos\theta)$ 
 and  ${\bi e}_2= (\cos\theta,\sin\theta)$ in the $x$-$y$ plane.  
In Fig.\ref{aniso}, we display $\tan\theta= -{ e}_{1y}/e_{1x}$, 
 $1-a_1/a_2$, and the $xy$ 
 component of the alignment tensor $Q_{xy}$ in (3.13).
All these quantities represent the degree of deformations of  
chain conformations in shear flow.  They 
 are insensitive to $T$ if plotted vs $\gdot\tau_{\rm R}$. 
For $\gdot\tau_{\rm R}\ls 1$, $\tan\theta$ is close to $1$ 
($\theta\cong  45^\circ$) and both 
$1-a_1/a_2$ and $Q_{xy}$ linearly increase with increasing
$\gdot\tau_{\rm R}$.
For $\gdot\tau_{\rm R}\gs 1$, these quantities 
saturate into   limiting values. At $T=0.2$, they are 
\be 
\theta\cong  80^\circ, \quad a_1/a_2 \cong 0.3, \quad 
Q_{xy}\cong 0.1.  
\label{4.7}
\en  
These are consistent with  $Q_{xy}\sim \sin\theta\cos\theta$.

\section{Transient viscoelastic behavior}
\setcounter{equation}{0}

In Fig.\ref{visgrowth},   
we plot the viscosity growth function
$\sigma_{xy}(t)/\gdot$ after application of shear $\gdot$ 
 at $t=0$ for $T=0.2$.  The system was at rest  for $t<0$.
In the initial stage $\gdot t \ls 0.1$, it evolves following  
the linear viscoelastic growth, 
\begin{equation}
\frac{1}{\gdot} \sigma_{xy}(t)  \cong 
\int_0^t dt' G(t').
\label{5.1}
\end{equation}  
In the nonlinear regime,   
$\sigma_{xy}(t)/\gdot$ tends to 
 the non-Newtonian  viscosity $\eta(\gdot)$.   
As a guide, we   also display the linear growth function 
$\int^t_0 dt' G_{\rm R}(t')$ in  the Rouse model.
In the  very early 
time region $1 \ll t \ls \tau_\alpha$, 
the growth ($\cong  G_0t$) 
is much larger  than  
the Rouse initial growth ($\cong  k_{\rm B}Tt$).  
We can see a rounded peak  for $\gdot=0.1$ before 
approach to the steady state  in Fig.\ref{vis}. 
More pronounced overshoot was already reported  at high shear 
in MD simulation of much longer 
unentangled alkane chains (C$_{100}$H$_{202}$).\cite{Cummings}

In experiments,\cite{muller,muller1,inoue,matsuyama}  
the stress-optical relation 
is transiently violated at low  $T$   after application of  elongational 
  flow due to the enhancement of the glassy component of 
 the  stress.   For shear flow,  Fig.\ref{sor} displays  our 
 MD results  at $T=0.2$ after application of shear 
 at $t=0$ in a stress-optical diagram,
where the solid lines are the  averages  over 
ten independent  runs.
As time goes on, the system traces the curve of 
a given $\gdot$, passes across 
the dashed curve representing  
the steady-state universal relation in Fig.\ref{steadySO}, 
and finally comes back to the dashed curve.
The deviation from the steady-state curve becomes  larger 
 with increasing $\gdot$,  as in 
the experiments of elongational flow.

\setcounter{equation}{0}
\section{Time-correlation functions 
and  tumbling  in shear flow }

In Fig.\ref{pps}, we show 
the end-to-end vector correlation functions 
$C(t)= \av{{\bi P}(t)\cdot{\bi P}(0)}/
\av{|{\bi P}|^2}$ with and without shear flow for 
various $\gdot$ at $T=0.2$. 
For $\gdot\ne0$, it rapidly decreases,  negatively 
overshoots, and finally  approaches  
zero with dumped oscillation superimposed.  
This   oscillatory behavior arises from 
random rotation of chains  in  shear flow, which is 
well known in dilute polymer solutions\cite{Mason,Chu,Wirtz} 
but has not been reported in polymer melts.
This is more evidently seen in Fig.\ref{rot}, where 
we show time development of 
the $x$ component of the end-to-end vector 
${\bi P}_j={\bi R}_N -{\bi R}_1$ of one  chain 
for $\gdot=10^{-3}$ (a), $10^{-2}$ (b), 
and $10^{-1}$ (c) at $T=0.2$. 
The corresponding Weisenberg number (4.1) is given by 
$Wi=  60, 600,$ and $6000$, respectively. In Fig.\ref{rot}(d), we show 
chain contours projected onto  the $x$-$y$ plane 
at the points $1 \sim 8$ indicated in Fig.\ref{rot}(c). 
When the chains 
change their  orientation, 
their shapes are contracted 
as in the case of a single  chain 
in solution.\cite{Chu} 
The average period of tumbling is about $35/\gdot$ 
in our case.

We may introduce  the van Hove time-correlation function (2.14) 
even  in shear flow if the particle 
displacement vector is redefined as\cite{yo2}   
\bea
\Delta {\bi R}_j(t) &=& 
{\bi R}_j(t+t_0)-{\bi R}_j(t_0) \nonumber\\
&-&\gdot\int_{0}^{t} dt'Y_{\rm G}(t_0+t'){\bi e}_x, 
\label{eq:6.1}
\ena
where $Y_{\rm G}$ is 
the $y$-component of ${\bi R}_{\rm G}= N^{-1}\sum_{j=1}^N{\bi R}_j$.
From the net displacement, the first term, 
 we have subtracted 
the flow-induced displacement, the second term. 
Figure \ref{fskts} 
shows $F_q(t)$ with $q=2\pi$  for various $\gdot$  at $T=0.2$.
Comparison of this figure with Fig.\ref{fsqt} suggests that
applying shear is analogous  to raising the temperature.
This tendency 
was already reported for the case of 
 supercooled binary mixtures.\cite{yo2,berthier}

We introduce the shear-dependent Rouse 
time $\tau_{\rm R}= \tau_{\rm R}(T,\gdot)$ and 
the $\alpha$ relaxation time 
$\tau_\alpha= \tau_\alpha(T,\gdot)$ by 
\be 
C(\tau_{\rm R})=e^{-1}, \quad 
F_q(\tau_\alpha)=e^{-1}.
\label{eq:6.2} 
\en 
We may then  examine how shear can accelerate the motions of 
chains and individual beads in shear flow. 
Figure \ref{taus} shows $\tau_{\rm R}$ and $\tau_\alpha$ 
 as functions of $\gdot$ at $T=0.2$.
In our short chain system, both 
 $\tau_{\rm R}$ and $\tau_\alpha$  decrease
for $\gdot\gs \tau_{\rm R}(T,0)^{-1} 
\ll\tau_\alpha(T,0)^{-1}$. 
Our data at $T=0.2$  are consistent with 
\be
\tau_{\rm R}(T,\gdot)^{-1}=
\tau_{\rm R}(T,0)^{-1}[1+A_{\rm R}\gdot],
\label{eq:6.3}
\en 
\be
\tau_\alpha(T,\gdot)^{-1}
=\tau_\alpha(T,0)^{-1}[1+(A_\alpha\gdot)^\mu],
\label{eq:6.4}
\en
where $A_{\rm R}\cong 10^4 \sim \tau_R(T,0)$, 
$A_\alpha \cong 6000\simeq 20\tau_\alpha(T,0)$, and $\mu \cong  0.77$. 
The average tumbling period  in Fig.12 
is about $4 \tau_{\rm R}(T,\gdot)$.  
For simple supercooled liquids we  already 
introduced  the van Hove 
time-correlation function   in shear flow\cite{yo2} and obtained  
$\tau_\alpha (T,\gdot)^{-1}
=\tau_\alpha(T,0)^{-1}[1+A_\alpha 
\gdot$] with $A_\alpha \sim \tau_\alpha(T,0)$.

The  sensitive shear-dependence of 
$\tau_\alpha(T,\gdot)$ predicted by (6.4) 
suggests potential importance of 
dielectric measurements in shear flow.\cite{yo2}
As a first experiment, Matsuyama {\it et al}. measured 
the dielectric  loss function $\epsilon''(\omega,\gdot)$ 
 in  steady shear $\gdot$  in 
 oligostyrene and polyisoprene melts.\cite{matsuyama}
In the former melt at $T=42^\circ$C, 
$\epsilon''(\omega,\gdot)$  decreased 
nonlinearly as a  function of  $\gdot$ 
at low frequencies $(\omega \ls 10^5$ s$^{-1}$) 
in the shear-thinning  regime $(\gdot \gs 15$ s$^{-1}$). 
Their  finding   indicates 
that $\tau_\alpha$ decreases 
as a function of $\gdot$  in the non-Newtonian 
regime, consistently with  (6.4), 
More systematic dielectric measurements in 
supercooled  systems under shear flow are 
very informative.

\setcounter{equation}{0}
\section{Summary}

We have performed very long MD  simulations
of  a supercooled 
polymer melt composed of $M=100$ 
short chains with bead number $N=10$ in quiescent and 
sheared conditions.  We here summarize our 
 main simulation results together with remarks.

(i) The stress relaxation function $G(t)$ is shown to 
follow  a stretched exponential decay (1.1) on the scale of 
the $\alpha$ relaxation time $\tau_\alpha$ and 
then the Rouse relaxation (3.6) 
on the scale of $\tau_{\rm R}$.

(ii) The nonlinear shear regime 
sets in at extremely small shear rate of order $\tau_{\rm R}^{-1}$ 
in supercooled states, where marked 
 shear-thinning and shape changes of chains
 are found.   Scattering and birefringence 
 experiments   from  weakly sheared melts near $T_{\rm g}$ 
seem to be very  promising.

(iii) In the nonlinear shear regime, each chain 
undergoes random tumbling in our melt as in the case of 
isolated polymer chains in shear flow. 
It is of great interest how this effect is universal 
in solutions and melts and how it influences macroscopic 
rheological properties. For example, 
we are interested in whether or not such tumbling occurs 
in sheared entangled polymers.

(iv)Transient stress divided by 
$\gdot$ after application of shear flow 
obeys  the 
linear  growth $\int_0^t dt'G(t')\simeq G_0 t$ for strain less 
than 0.1 and then 
saturates into a non-Newtonian steady-state viscosity.
This initial growth is much steeper than that predicted by 
the Rouse model. As a result,  
the stress-optical relation does not hold 
transiently under deformation
even in the linear (zero-shear) limit.
Its violation is more enhanced for larger shear 
rates. These are consistent 
with the experiments.

(v) The time-correlation functions
in shear flow are calculated for the end-to-end vector 
and the modified particle displacement in (6.1). 
The former represents the relaxation of chain conformations, 
while the latter the monomeric relaxation 
on the spatial scale of the particle distance. 
We can then determine the shear-dependent 
relaxation times, 
 $\tau_{\rm R} (T,\gdot)$ 
and  $\tau_{\alpha} (T,\gdot)$.
They decrease nonlinearly and behave differently 
as functions of $\gdot$ in the nonlinear shear regime 
as in (6.3) and (6.4). 
It is then of great interest how 
these times behave in strong shear for much larger 
$N$. We conjecture that 
if $N$ is sufficiently large, 
shear should first influence the overall 
chain conformations, while it does not much 
affect the monomeric relaxations. 
We propose dielectric measurements 
in shear flow, which should give information of 
$\tau_{\alpha} (T,\gdot)$.\cite{matsuyama}  
In addition, as demonstrated in Fig.13, the effect of shear 
on the van Hove self-correlation function is analogous to 
raising the temperature above $T_{\rm g}$ as in supercooled 
binary mixtures.\cite{yo2,berthier}

\section*{Acknowledgments}
We would like to thank 
 T. Inoue for valuable discussions on the stress-optical 
 relation. This work is supported by Grants in Aid for Scientific 
Research from the Ministry of Education, Science, Sports and Culture
of Japan.
Computations have been performed 
at the Human Genome Center, Institute of Medical Science, University 
of Tokyo.

\end{multicols}

\narrowtext
\begin{figure}[t]
\epsfxsize=2.8in
\centerline{\epsfbox{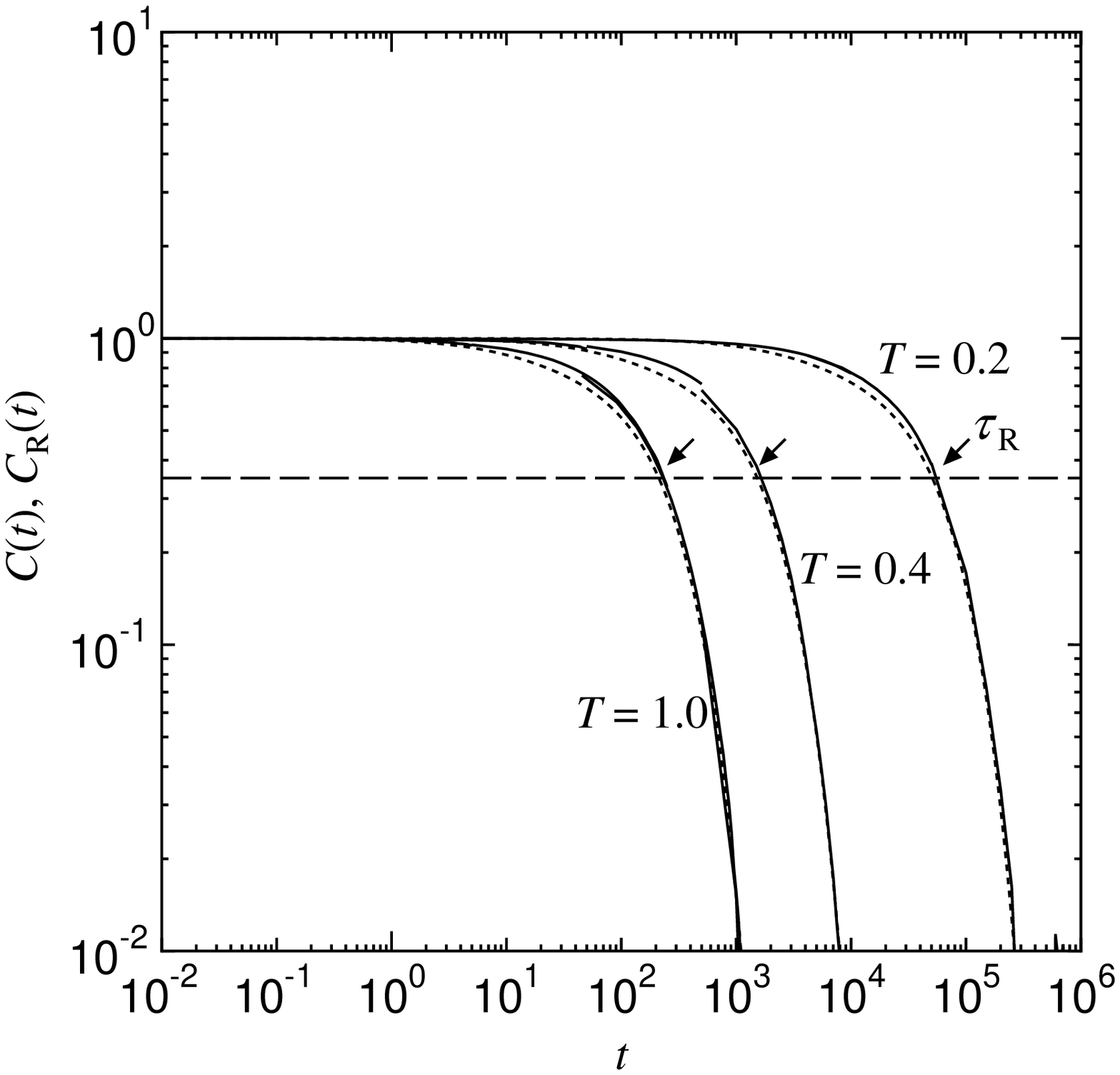}}
\caption{\protect
Normalized  end-to-end vector time-correlation function 
$C(t)$ in (3.4) for  $T=1.0$, $0.4$, and $0.2$.  
The dotted lines are the results 
of the Rouse model (3.5). The Rouse time $\tau_{\rm R}$ 
in (3.2) is indicated by arrows. 
}
\label{pp}
\end{figure}

\begin{figure}[b]
\epsfxsize=2.8in
\centerline{\epsfbox{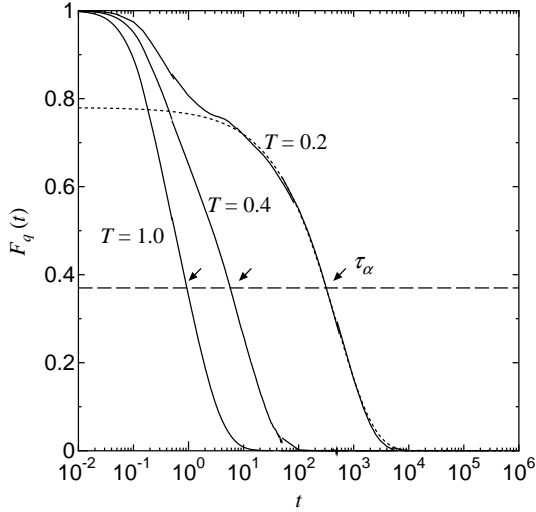}}
\caption{\protect
The van Hove self-correlation function 
$F_q(t)$  at $q=2\pi$ for 
 $T=1.0$, $0.4$, and $0.2$ on a semi-logarithmic scale.  
The dotted line represents the stretched 
exponential decay  $\propto 
\exp [-(t/\tau_\alpha)^{0.64}]$. }
\label{fsqt}
\end{figure}

\begin{figure}[t]
\epsfxsize=2.8in
\centerline{\epsfbox{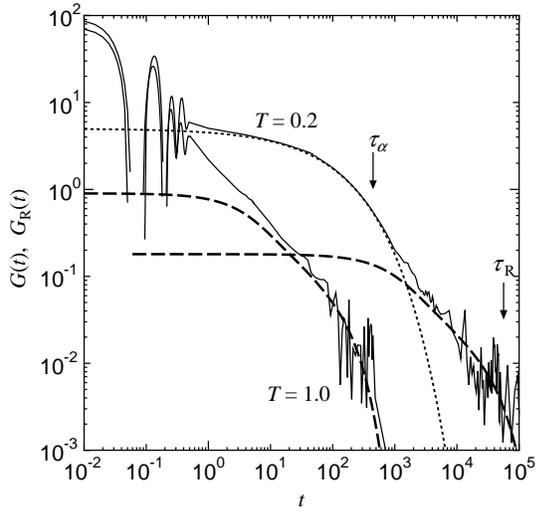}}
\caption{\protect
The  stress relaxation function $G(t)$ in (3.10) 
(thin-solid lines) at $T=0.2$ in a supercooled state  and 
$T=1$ in a normal liquid state. 
For $T=0.2$, it can be  fitted to the stretched exponential form,  
 $\exp [-(t/\tau_{\rm s})^{0.5}]$ with $\tau_{\rm s}=90$, 
(dotted line) for $1 \ls t \ls 10^3$ 
and  tends to the Rouse relaxation function 
$G_{\rm R}(t)$ (bold-dashed lines) at later times.
}
\label{gt}
\end{figure}

\begin{figure}[t]
\epsfxsize=2.8in
\centerline{\epsfbox{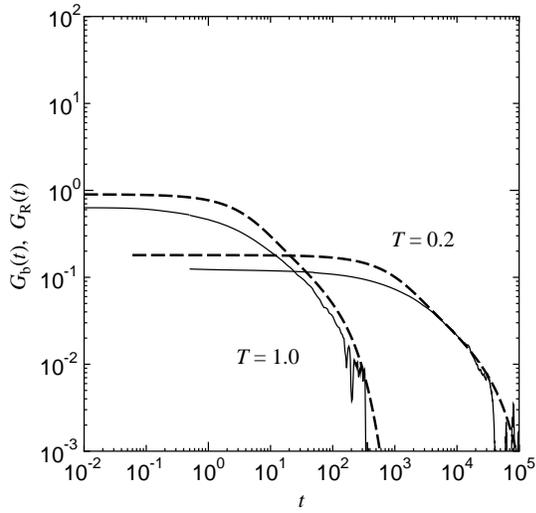}}
\caption{\protect
Comparison of 
the time-correlation  function $G_{\rm b}(t)$ in (3.14)  
(thin-solid lines) and 
the Rouse relaxation function  $G_{\rm R}(t)$ 
in (3.6) (bold-dashed lines).
}
\label{grt}
\end{figure}

\begin{figure}[t]
\epsfxsize=2.8in
\centerline{\epsfbox{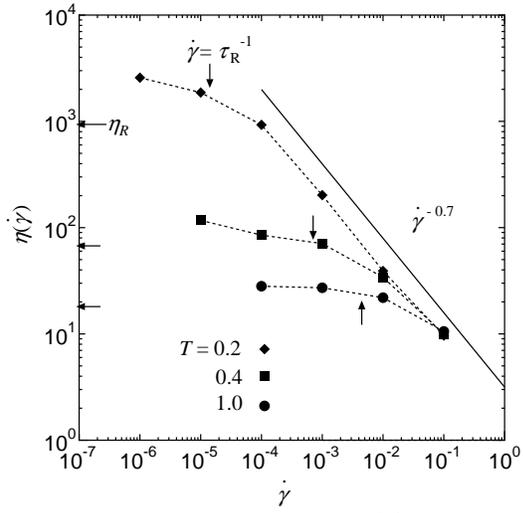}}
\caption{\protect
The  steady-state viscosity $\eta(\gdot)$ vs 
shear $\gdot$ 
for $T=0.2,0.4,$ and $1$. A line of slope $-0.7$ is 
 a view guide.
}
\label{vis}
\end{figure}

\begin{figure}[t]
\centerline{\epsfxsize=2.8in\epsfbox{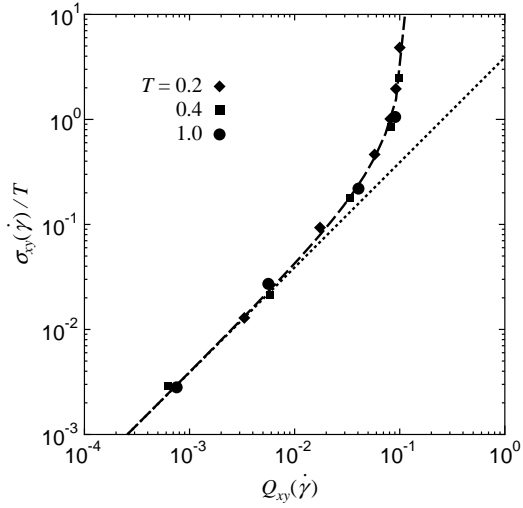}}
\caption{\protect
Universal stress-optical relation 
$\sigma_{xy}/T$ vs $Q_{xy}$ 
in steady states under shear flow for 
$T=0.2, 0.4,$ and $1$. 
}
\label{steadySO}
\end{figure}

\widetext
\begin{figure}[t]
\centerline{
\epsfxsize=6.5in\epsfbox{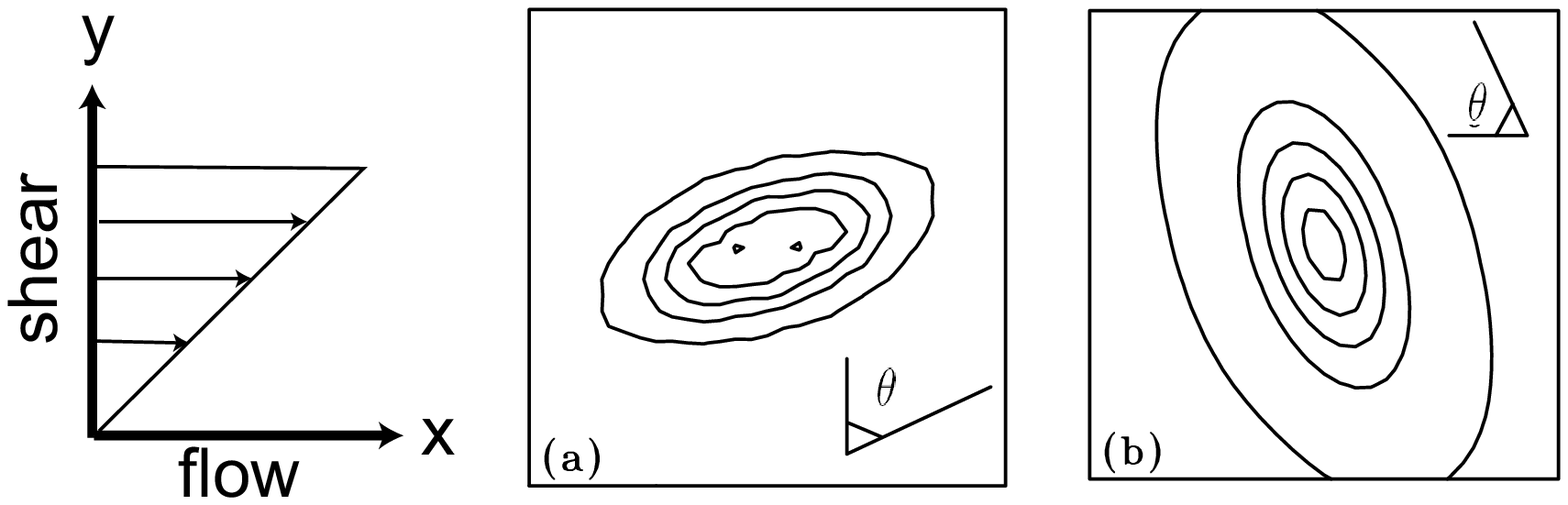}}
\caption{\protect
(a) Isointensity curves of  $ g_{\rm s}({\bf r})$ in (4.3) 
in the $x$-$y$ plane ($-3.75<x,y<3.75,~ z=0$). 
(b) Those  of the incoherent structure factor 
$S({\bf q})$ in (4.4) 
in the $q_x$-$q_y$ plane ($-\pi<q_x,q_y<\pi, ~q_z=0$).  
The values on the isolines are $0.01+0.02n$ in (a) 
and $0.1+0.2n$ in (b) with $n=0,1,\cdots,4$ from outer to inner.
Here  $T=0.2$,  $\gdot=10^{-4}$, 
and the  flow is  in the horizontal ($x$) direction.
The $\theta$ is the angle between  the average 
chain shapes and  the $y$ axis. 
}
\label{ellips}
\end{figure}

\narrowtext
\begin{figure}[t]
\epsfxsize=2.8in
\centerline{\epsfbox{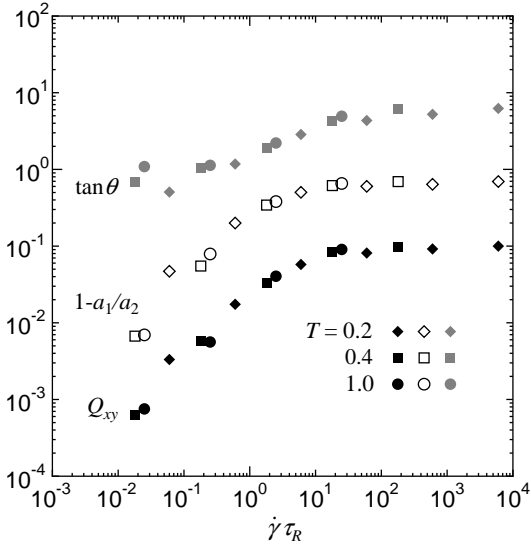}}
\caption{\protect
$\tan\theta$, $1-a_1/a_2$, and $Q_{xy}$ vs 
$\gdot \tau_{\rm R}$ in steady states.
}
\label{aniso}
\end{figure}

\begin{figure}[b]
\epsfxsize=2.8in
\centerline{\epsfbox{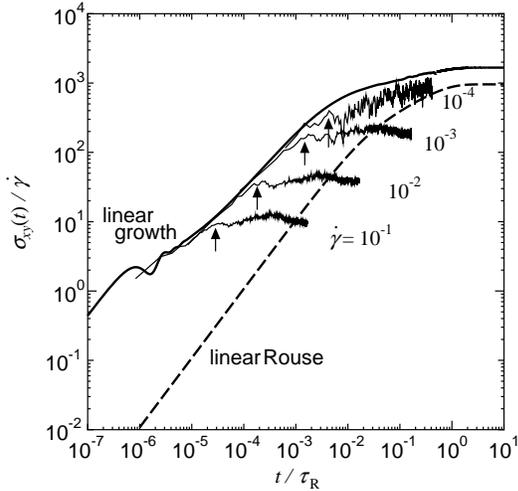}}
\caption{\protect
Shear stress  divided by shear rate 
$\sigma_{xy}(t)/\gdot$ 
vs $t/\tau_{\rm R}$ 
   for  $\gdot=10^{-1}$, $10^{-2}$, $10^{-3}$, $10^{-4}$ 
(thin-solid lines)   
at  $T=0.2$ where $\tau_{\rm R}=6\times 10^4$.  
The curves  follow   the linear viscosity growth 
function (bold-solid line) 
for $\gdot t \ls 0.1$,  
but  depart  from it for  $\gdot t \gs 0.1$.  
The linear growth function in the Rouse model 
is also plotted(bold-dashed line).  The arrows indicate 
onset of the nonlinear behavior. 
}
\label{visgrowth}
\end{figure}

\begin{figure}[t]
\epsfxsize=2.8in
\centerline{\epsfbox{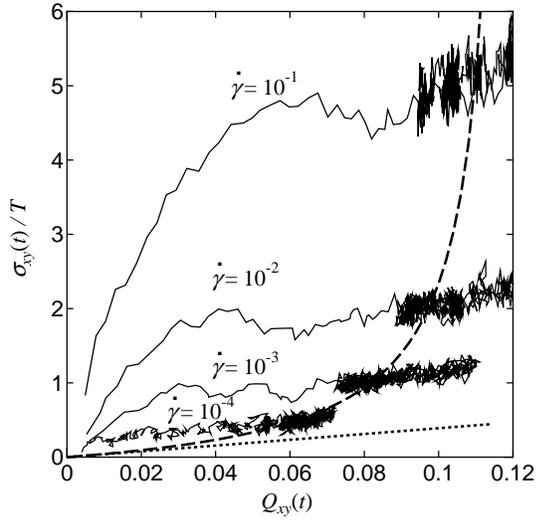}}
\caption{\protect
Parametric plots of $\sigma_{xy}(t)/T$ vs  
$Q_{xy}(t)$ after application of shear  at $t=0$ 
for $T=0.2$. The curves initially  deviate 
from  the universal steady-state curve  
obtained in Fig.\ref{steadySO} (dashed line) but 
 approach it ultimately.  The deviations increase with increasing $\gdot$.
}
\label{sor}
\end{figure}

\begin{figure}[t]
\epsfxsize=2.8in
\centerline{\epsfbox{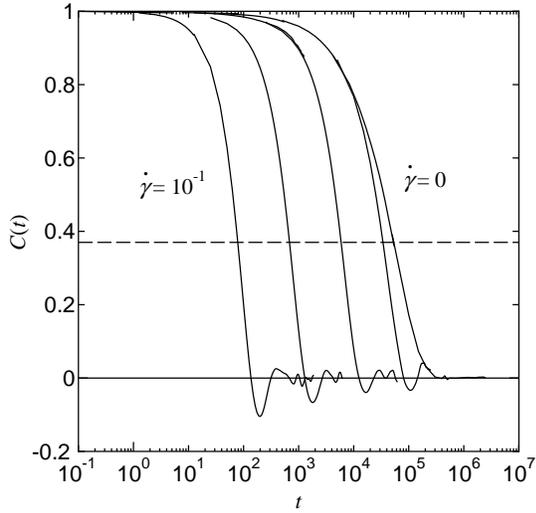}}
\caption{\protect
Normalized  time-correlation function 
of the end-to-end vector $C(t)$ in (3.4)
 at $T=0.2$ for $\gdot=0$, 
$10^{-4}$, $10^{-3}$, $10^{-2}$, and $10^{-1}$ from right to left
on a semi-logarithmic scale.
The negative overshoot for 
$\gdot > 0$ arises from rotational motions
of chains.
}
\label{pps}
\end{figure}

\widetext
\begin{figure}[t]
\epsfxsize=5.5in
\centerline{\epsfbox{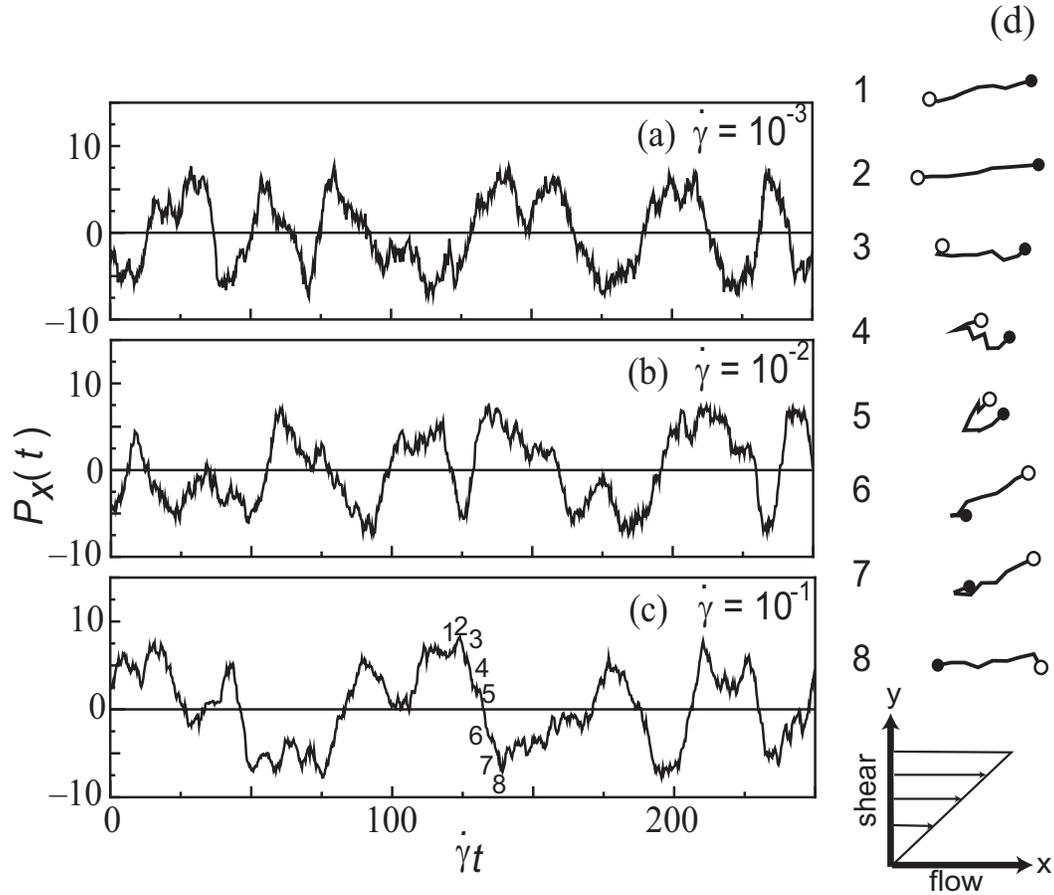}}
\caption{\protect
Time-evolution of the $x$ component of 
the end-to-end vector 
$P_x(t)={X}_N(t) - X_1(t)$ 
of one chain vs $\gdot t$. 
Here  $T=0.2$ and  $\gdot=10^{-3}$(a), $10^{-2}$(b), and $10^{-1}$(c)  
from above.  
Typical tumbling motions at the points $1 \sim 8$ indicated in (c)
are shown in (d), where the chain conformations 
are projected on the $x$-$y$ plane.
}
\label{rot}
\end{figure}


\narrowtext
\begin{figure}[t]
\epsfxsize=2.8in
\centerline{\epsfbox{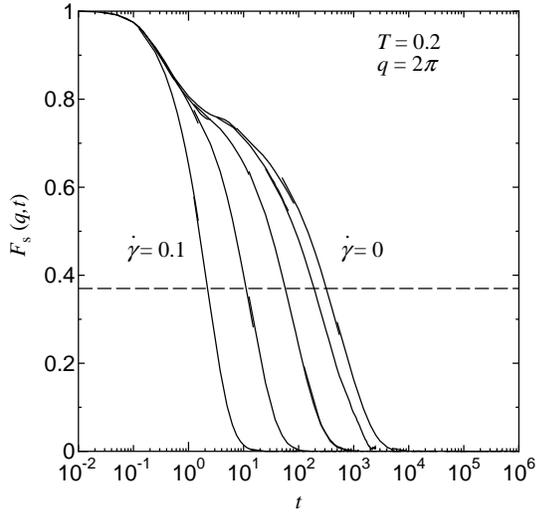}}
\caption{\protect
The van Hove self correlation function (3.7) 
with (6.1) at $T=0.2$ for  $\gdot=0$, 
$10^{-4}$, $10^{-3}$, $10^{-2}$, and $10^{-1}$ from  right to left
 on a semi-logarithmic scale.
}
\label{fskts}
\end{figure}

\begin{figure}[t]
\epsfxsize=2.8in
\centerline{\epsfbox{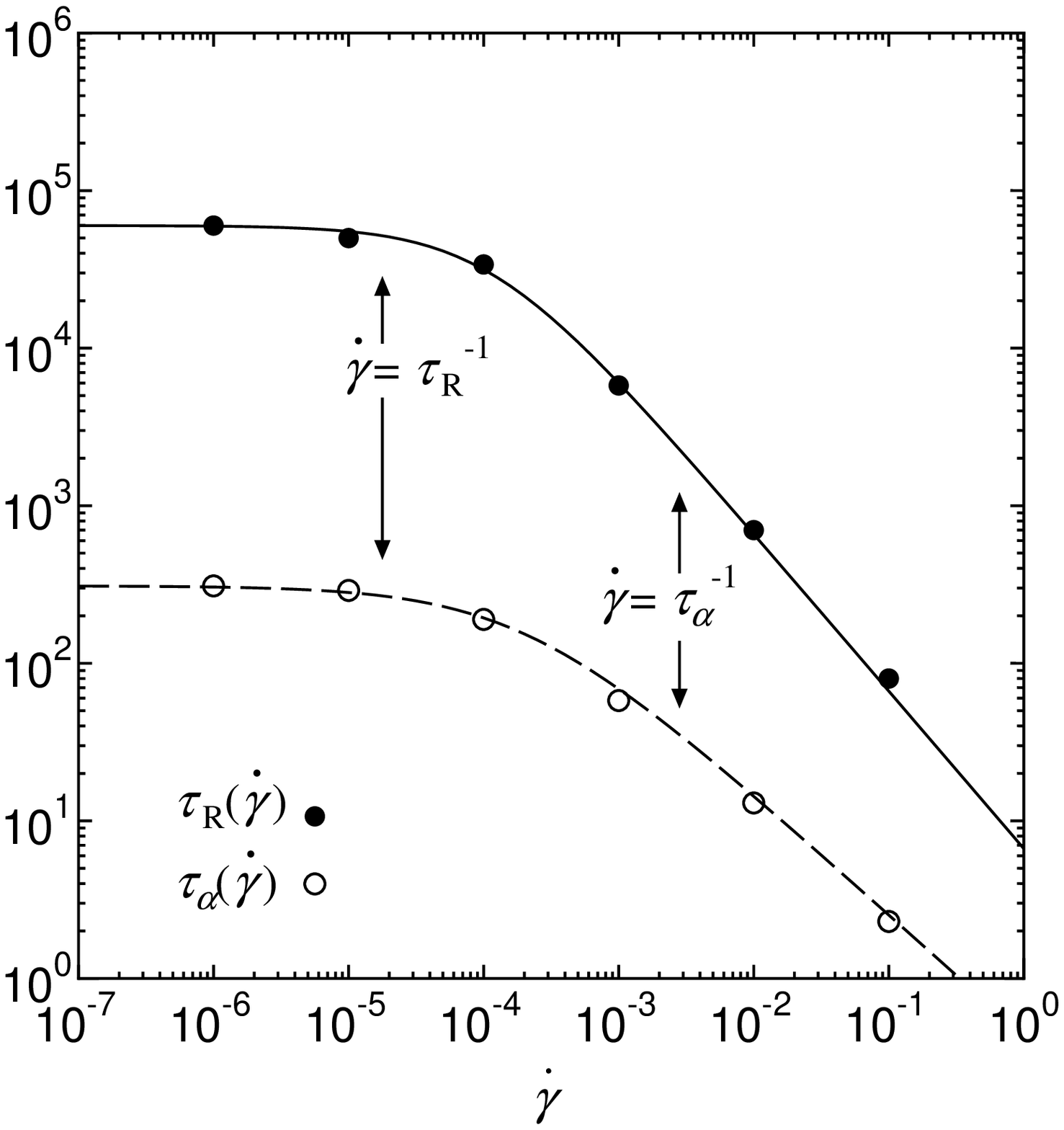}}
\caption{\protect
Two relaxation times  $\tau_{\rm R}(\gdot)$ 
and $\tau_\alpha(\gdot)$ as functions of 
shear $\gdot$ at $T=0.2$ determined from  (6.2).
Both these times decrease  for 
$\gdot \gs
\tau_{\rm R}(0)^{-1}\sim  N^{-2}\tau_\alpha(0)^{-1}$ in our 
short chain system.  
The solid and dashed lines represent (6.3) and (6.4), respectively.
The slopes of the curves at high shear 
are $-1$ for $\tau_{\rm R}(\gdot)$  and $-0.77$ for 
$\tau_\alpha(\gdot)$. 
}
\label{taus}
\end{figure}


\begin{references}


\bibitem{matsuoka}
S. Matsuoka,
{\it Relaxation Phenomena in Polymers}, 
(Oxford, New York, 1992).

\bibitem{stroble} G.R. Stroble, {\it The Physics of Polymers}, 
(Springer, Heidelberg, 1996). 

\bibitem{doi}
M. Doi and S.F. Edwards, 
{\it The Theory of Polymer Dynamics}
(Clarendon, Oxford, 1986).

\bibitem{Simmons}  J.H. Simmons, R. Ochoa, K.D. Simmons and J.J. Mills, 
{J. Non-Cryst. Solids} {\bf 105}, 313 (1988).

\bibitem{yo1} R. Yamamoto and A. Onuki, 
Europhys. Lett. {\bf 40}, 61 (1997). 

\bibitem{yo2} R. Yamamoto and A. Onuki, 
Phys. Rev. E {\bf 58}, 3515 (1998). 

\bibitem{yo3} R. Yamamoto and A. Onuki, 
J. Phys.; Condens. Matter {\bf 29}, 6323  (2000).


\bibitem{KLH}
M. Kr\"oger, W. Loose, and S. Hess,
J. Rheology {\bf 37}, 1057 (1993).


\bibitem{ch}
S. Chynoweth and Y. Michopoulos,
Molec. Phys. {\bf 81}, 133 (1994).

\bibitem{khare}
R. Khare and J. de Pablo,
J. Chem. Phys. {\bf 107}, 6956 (1997).

\bibitem{Moore} J.D. Moore, S.T. Cui, H.D. Cochran, and P.T. Cummings,  
J. Non-Newtonian Fluid Mech. {\bf 93}, 83 (2000); {\bf 93}, 101 (2000).

\bibitem{Cummings} S.  Bair, C. McCabe, and P.T. Cummings,  
Phys. Rev. Lett. {\bf 88}, 058302 (2002).

\bibitem{Ara1} P.H. Mott,  A.S. Aragon  and 
U.W. Suter, Philos. Mag. A,  {\bi 67}, 931 (1993). 
\bibitem{Ara2}  A.S. Aragon, 
V.V. Bulatov, P.H. Mott and U.W. Suter, J. Rheol.  {\bi 39}, 377
(1995).



\bibitem{onuki-doi} 
A.\ Onuki and M.\ Doi, J.\ Chem.\ Phys.\ {\bf 85}, 1190 (1986).


\bibitem{muller}
R. Muller and J.J. Pesce,
Polymer {\bf 35}, 734 (1994). 

\bibitem{muller1} 
M. Kr\"oger, C. Luap, and R. Muller,
Macromolecules {\bf 30}, 526 (1997).

\bibitem{inoue}
T. Inoue, D.S. Ryu, and K. Osaki,
Macromolecules {\bf 31}, 6977 (1998).

\bibitem{matsuyama}
M. Matsuyama, H. Watanabe, T. Inoue  ,and K. Osaki,
Macromolecules   {\bf 31}, 7973 (1998).  

\bibitem{kremer}
K. Kremer and G.S. Grest,
J. Chem. Phys. {\bf 92}, 5057 (1990).
\bibitem{carmesin}
I. Carmesin and K. Kremer,
Macromolecules {\bf21}, 2819 (1988).
\bibitem{deutsch}
H.P. Deutsch and K. Binder,
J. Chem. Phys. {\bf 94}, 2249 (1991).
\bibitem{Roe}
D. Rigby and R.-J. Roe,
J. Chem. Phys. {\bf 87}, 7285 (1987);
R.-J. Roe,
J. Chem. Phys. {\bf 100}, 1610 (1994).

\bibitem{bennemann}
C. Bennemann, W. Paul, K. Binder, and B. D\"unweg,
Phys. Rev. E {\bf 57}, 843 (1998);
C. Bennemann, J. Baschnagel, W. Paul,
Eur. Phys. J. B {\bf 10}, 323 (1999);
C. Bennemann, J. Baschnagel, W. Paul, and K. Binder,
Comp. Theo. Poly. Sci. {\bf 9}, 217 (1999).

\bibitem{baschnagel}
J. Baschnagel, C. Bennemann, W. Paul, and K. Binder,
J. Phys.; Condens. Matter {\bf 12}, 6365 (2000).


\bibitem{moe}
N.E. Moe and M.D. Ediger,
Phys. Rev. E {\bf 59}, 623 (1999).

\bibitem{zon}
A. van Zon and S.W. de Leeuw,
Phys. Rev. E {\bf 60}, 6942 (1999).

\bibitem{okun} K. Okun,  M. Wolfgardt, 
J. Baschnagel, and  K. Binder, 
Macromolecules {\bf 30}, 3075 (1997).  

\bibitem{paul}
W. Paul and J. Baschnagel, 
in {\it Monte Carlo and Molecular Dynamics Simulations in Polymer
Science},
K. Binder ed. 
(Oxford University Press, New York, 1995), pp. 307-355. 

\bibitem{binder} K. Binder, C. Bennemann, J. Baschnagel and W. Paul,
in {\it Anomalous diffusion}, 
R. Kutner, A. Pekalski, and K. Sznajd-Weron, eds., 
(Springer, Berlin, 1999), pp. 124-139;
K. Binder, J. Baschnagel, C. Bennemann, and W. Paul,
J. Phys.: Condens. Matter {\bf 11}, A47-A55 (1999).

\bibitem{rouault}
Y. Rouault and K. Kremer
J. Chem. Phys. {\bf 111}, 3288 (1999).

\bibitem{Onukibook} A. Onuki, {\it Phase Transition Dynamics}, 
(Cambridge University Press, Cambridge, 2002).


\bibitem{Allen}  M.P. Allen  and  D.J. Tildesley,
{\it Computer Simulation of Liquids}
(Clarendon, Oxford, 1987).

\bibitem{Evans}  D.J. Evans and G.P. Morriss,
{\it  Statistical Mechanics of Nonequilibrium Liquids}
(Academic, New York, 1990).

\bibitem{kopf}
A. Kopf, B. D\"unweg, and W. Paul,
J. Chem. Phys. {\bf 107}, 6945 (1997).

\bibitem{verdier}
P.H. Verdier,
J. Chem. Phys. {\bf 45}, 2118 (1966).



\bibitem{picot}
R. Muller, J.J. Pesce, and C. Picot,
Macromolecules {\bf 26}, 4356 (1993).


\bibitem{Mason} I.Y.Z. Zia, R.G. Cox and S.G. Mason, 
Proc. Roy. London A {\bf 300}, 421 (1967). 
\bibitem{Chu} D.E. Smith, H.P. Babcock and S. Chu, 
Science {\bf 283}, 1724 (1999). 
\bibitem{Wirtz} P. LeDuc, C. Haber, G. Bao 
and D. Wirtz, Nature {\bf 399}, 564 (1999). 


\bibitem{berthier}
L. Berthier and J.-L. Barrat,
Phys. Rev. E {\bf 63}, 012503 (2001);
J. Chem. Phys. in print.


\end{references}
\end{document}